\documentclass[12pt,dvips]{article}
\usepackage{graphicx}
\usepackage{amsmath}
\usepackage[psamsfonts]{amssymb}
\usepackage{amsthm}
\usepackage{euscript}

\usepackage{latexsym}

\renewcommand{\theequation}{\arabic{section}.\arabic{equation}}
\renewcommand{\(}{\begin{equation}}
\renewcommand{\)}{end{equation} \vspace{-.05in}\linebreak}

\newcounter{saveeqn}
\newcounter{savealpheqn}

\newcommand{\alpheqn}{\setcounter{saveeqn}{\value{equation}}%
 \stepcounter{saveeqn}\setcounter{equation}{0}%
 \renewcommand{\theequation}{\mbox{\arabic{section}.\arabic{saveeqn}\alph{equation}}}
 \renewcommand{\)}{\end{equation}}}
\def\part#1{\frac{\partial}{\partial{#1}}}%
\def\group#1{\refstepcounter{equation}\setcounter{saveeqn}{\value{equation}}%
 \label{#1}\setcounter{equation}{0}%
\renewcommand{\theequation}{\mbox{\arabic{section}.\arabic{saveeqn}\alph{equation}}}
 \renewcommand{\)}{\end{equation}}}
\newcommand{\reseteqn}{\setcounter{equation}{\value{saveeqn}}%
 \renewcommand{\theequation}{\arabic{section}.\arabic{equation}}%
 \renewcommand{\)}{\end{equation}}}

\newcommand{\aalpheqn}{\setcounter{saveeqn}{\value{equation}}%
 \stepcounter{saveeqn}\setcounter{equation}{0}%
 \renewcommand{\theequation}{\mbox{\Alph{subsection}.\arabic{saveeqn}\alph{equation}}}
  \renewcommand{\)}{\end{equation}}}
\newcommand{\areseteqn}{\setcounter{equation}{\value{saveeqn}}%
 \renewcommand{\theequation}{\Alph{subsection}.\arabic{equation}}%
 \renewcommand{\)}{\end{equation}}}

\renewcommand{\thefootnote}{\alph{footnote}}
\renewcommand{\(}{\begin{equation}}
\renewcommand{\)}{\end{equation}}
\newcommand{\ba}{\begin{eqnarray}}
\newcommand{\ea}{\end{eqnarray}}

\newcommand{\bp}{\mathop{\vtop{\ialign{##\crcr
  $\hfil\displaystyle{}\hfil$\crcr\noalign{\kern-13pt\nointerlineskip}
  \BIG{(}\hskip0pt\crcr\noalign{\kern3pt}}}}}
\newcommand{\cbp}{\mathop{\vtop{\ialign{##\crcr
  $\hfil\displaystyle{}\hfil$\crcr\noalign{\kern-13pt\nointerlineskip}
  \BIG{)}\hskip0pt\crcr\noalign{\kern3pt}}}}}
\newcommand{\pa}{\mathop{\vtop{\ialign{##\crcr
  $\hfil\displaystyle{\oplus}\hfil$\crcr\noalign{\kern+1pt\nointerlineskip}
  \hspace{.08in}$^{\alpha=0}$\hskip6pt\crcr\noalign{\kern3pt}}}}}
\renewcommand{\sp}{,\hspace{.3in}}

\newcommand{\R}{\ensuremath{\mathbb R}}

\newcommand{\Z}{\ensuremath{\mathbb Z}}

\newcommand{\beq}{\begin{equation}}
\newcommand{\eeq}{\end{equation}}
\newcommand{\sub}{\subset}

\newcommand{\into}{\hookrightarrow}

\numberwithin{equation}{section}
\def\hsp#1{\hspace{#1in}}

\catcode`\@=11
\def\vereq#1#2{\lower3pt\vbox{\baselineskip1.5pt \lineskip1.5pt
\ialign{$\m@th#1\hfill##\hfil$\crcr#2\crcr\sim\crcr}}}
\catcode`\@=12

\makeatletter
\newcommand\figcaption{\def\@captype{figure}\caption}
\newcommand\tabcaption{\def\@captype{table}\caption}
\makeatother
\renewcommand{\(}{\begin{equation}}
\renewcommand{\)}{\end{equation}}

\begin{document}
\begin{titlepage}
\begin{flushright}
UCB-PTH-01/44 \\
hep-th/0112084
\end{flushright}

\vspace{2em}
\def\thefootnote{\fnsymbol{footnote}}

\begin{center}
{\Large K-Theory and S-Duality: Starting Over from Square 3}
\end{center}
\vspace{1em}

\begin{center}
Jarah Evslin\footnote{E-Mail: jarah@uclink4.berkeley.edu} and Uday Varadarajan\footnote{E-Mail: udayv@socrates.berkeley.edu} 
\end{center}

\begin{center}
\vspace{1em}
{\em Department of Mathematics,
     University of California\\
     Berkeley, California 94720}\\
{\hsp{.3}}\\
{\em Department of Physics,
     University of California\\
     Berkeley, California 94720}\\
\hsp{.3}\\
{\em Theoretical Physics Group\\
     Ernest Orlando Lawrence Berkeley National Laboratory\\
     1 Cyclotron Rd,
     Berkeley, California 94720}

\end{center}

\vspace{3em}
\begin{abstract}
\noindent
Recently Maldacena, Moore, and Seiberg (MMS) have proposed a physical
interpretation of the Atiyah-Hirzebruch spectral sequence, which
roughly computes the K-homology groups that classify D-branes. We note
that in IIB string theory, this approach can be generalized to
include NS charged objects and conjecture an S-duality covariant,
nonlinear extension of the spectral sequence.  We then compute the
contribution of the MMS double-instanton configuration to the
derivation $d_5$.  We conclude with an M-theoretic generalization
reminiscent of 11-dimensional $E_8$ gauge theory.

\end{abstract}

\vfill
December 11, 2001

\end{titlepage}
\setcounter{footnote}{0} 
\renewcommand{\thefootnote}{\arabic{footnote}}

\pagebreak
\renewcommand{\thepage}{\arabic{page}}
\pagebreak 

\section{Introduction}
\group{dummy}\reseteqn
Various manifestations of K-theory appear to classify D-brane
configurations at large distances and weak coupling \cite{MM,WittenK}.
This is a consequence of the proposal \cite{Sen} that each
configuration is related, by the dynamical process of tachyon
condensation, to a universe filled with non-BPS D9-branes
\cite{HoravaIIA} or D9-$\overline{\textup{D}9}$ pairs whose tachyon
fields are sections of bundles over the 10-dimensional spacetime.  One
can then define the configurations before and after tachyon
condensation to be equivalent.  Tachyon configurations are classified
by K-theory \cite{WittenK} and so we learn that K-theory classifies
D-brane configurations as well.

In IIB string theory, S-duality would suggest a similar classification
of NS and RR charged objects. However, tachyon configurations on space
filling branes do not obviously yield NS5-branes or fundamental
strings\footnote{For an exception, see \cite{HarveyMoore}.}. Of
course, one could simply impose S-duality and then arrive at a
``K-theory'' classification that includes NS charged objects, but some
authors \cite{AspinwallPlesser} have suggested that perhaps S-duality
itself fails when discrete torsions are involved.

While the tachyon condensation approach to D-brane classification is
difficult to generalize to other branes, the approach recently
proposed by Maldacena, Moore, and Seiberg \cite{MMS} generalizes
beautifully.  Consider all of the consistent time independent states
and identify the ones related by physical processes, which are
manifested in examples as ``instantonic'' branes.  In the case of
D-branes in an H-field, this prescription may be used to construct an
Atiyah-Hirzebruch \cite{AH} spectral sequence which computes the
K-group of a space (modulo an extension problem related to the
dimension of torsional brane charges \cite{Gimon}). We introduce an
extension of this procedure which includes fundamental strings,
NS5-branes and branes in M-theory.

We will require three tools.  The non-torsion part of the problem will
be understood using the classical equations of motion and Bianchi
identities from IIB and 11-dimensional supergravity. In particular,
the nonlinearities in supergravity translate into corresponding
nonlinearities in the ``differentials'' of the sequence. Incorporating
torsion corrections in IIB will require a variation of the
Freed-Witten anomaly, which restricts NS5-branes to wrap spin$^c$
submanifolds of spacetime to avoid D-string worldsheet anomalies.  The
nontorsion classification in M-theory will be understood via an
interpretation of M5-branes as defects in an $E_8$ bundle over the
11-dimensional spacetime, the restriction to 11 dimensions of a
12-dimensional construction that appeared in
Refs.~\cite{Stong,FluxQuant,DMW}.

In section \ref{gensec}, after briefly reviewing the Atiyah-Hirzebruch
spectral sequence, we present our conjecture for an S-duality
covariant extension of the sequence.  Then in Sec.~\ref{sugsec} we
review the facts that we will need from classical IIB supergravity and
review branes ending on branes that wrap a cycle supporting flux.  In
the following section we present the MMS interpretation of the
spectral sequence and generalize it to include NS5-branes and strings
in the supergravity limit, where torsion is neglected.  In the next
section the Freed-Witten anomaly is reviewed and generalized, which
allows torsion to be incorporated in the new sequence.  In section
\ref{susec} examples are given and in particular we consider the SU(3)
WZW model in some detail.  The double-instanton found in
Ref.~\cite{MMS} is generalized and its contribution to the
differential $d_5$ is computed.  This analysis is extended to include
NS5-branes and strings in the SU(3) case.  In the last section the
techniques of this paper are applied to M theory.

\section{The Conjecture}\label{gensec}
\subsection{The AHSS} \label{origsec}
The Atiyah-Hirzebruch spectral sequence \cite{AH} is an algorithm which
relates cohomology and the K-groups $K^*(M)$. First introduce a filtration
on $K(M)$ by defining
\begin{equation}
K_p = \textup{Ker\ }K(M) \rightarrow K(M^p)
\end{equation}
where $M_p$ is the p-skeleton of M.  The spectral sequence then
computes, for example, the associated graded algebra of $K^1(M)$
\begin{equation}
\textup{Gr}K^1=\oplus_q K_{q+1}^1/K_{q}^1.
\end{equation}

This process procedes through a series of approximations $K^1\sim
E_n$ which terminate after a
finite number of iterations. The first approximation is integer-valued cohomology
\begin{equation}
E_1=\bigoplus_{j\textup{\ odd}}E_1^{j}=\bigoplus_{j\textup{\
    odd}}H^{j}(M,\Z). \label{e1}
\end{equation}
Successive approximations result from taking the cohomology of
(\ref{e1}) with respect to a sequence of differentials
\begin{equation}
d_{p+2}:E_{p}^q\rightarrow E_{p}^{q+p+2}
\end{equation}
where $p$ is odd.  That is,
\begin{equation}
E_{p+2}=\textup{ker}(d_{p+2})/\textup{Im}(d_{p+2})
\end{equation}
giving $E_{p+2}$ as an equivalence class of subsets of $E_{p}$.  For
sufficiently high $p$,
\begin{equation}
\textup{Gr}K^1=E_{p+2}
\end{equation}
and then $K^1$ can be computed by the solution of an extension problem.

In IIA string theory only $d_3$ is needed, that is, the associated
graded algebra is simply $E_3$.  In IIB this is also true with the
exception of $d_5$ acting on the 3 form fieldstrengths of 5-branes in
the presence of a nontrivial H flux, which is sometimes nontrivial
\cite{MMS}. As we will review later,
\begin{equation}
d_3=Sq^3+H
\end{equation}
and so
\begin{equation}
\textup{Gr}K^1=\textup{Ker}(Sq^3+H)/\textup{Im}(Sq^3+H) \label{oldk}
\end{equation}
up to $d_5$ corrections.  Thus (\ref{oldk}) classifies which integral
cohomology classes may be realized as RR fluxes in string theory.  Our
goal is to extend this formula to include NS fluxes in IIB.

\subsection{S-Duality Covariant AHSS}
We propose the following modified Atiyah-Hirzebruch spectral sequence
as a starting point for an S-duality covariant classification for
fluxes in IIB string theory. Instead of beginning with the complex of
odd dimensional cohomology classes (\ref{e1}), one begins with
\begin{equation}
E_1=H^1\bigoplus H^3\bigoplus H^3\bigoplus H^5\bigoplus H^7\bigoplus H^7. 
\end{equation}

As we will explain below, the MMS interpretation of the differentials
combined with a generalization of the Freed-Witten anomaly argument
suggest
\group{main}
\begin{equation}
d_3^1(G_1)=(Sq^3+H\cup)G_1
\end{equation}
\begin{equation}
d_3^3(G_3,H)=Sq^3(G_3+H)+G_3\cup H \label{conjb}
\end{equation}
\begin{equation}
d_3^{5a}(G_5)=(Sq^3+H\cup) G_5\sp
d_3^{5b}(G_5)=(Sq^3+G_3\cup) G_5
\end{equation}
\begin{equation}
d_3^7(*G_3,*H)=Sq^3(*G_3+*H)+H\cup *G_3.
\end{equation}
\reseteqn 
Notice that the $Sq^3$ terms are trivial in all but (\ref{conjb}).
We claim that flux configurations which are not annihilated by the
above differentials are anomalous.  In particular, in the supergravity
limit where one ignores the torsion terms, this condition is
equivalent to the enforcement of supergravity equations of motion, to
be reviewed in Subsec.~\ref{sugrasec}.  In addition we claim that flux
configurations that are exact under either of these differential
operators are unstable, and decay via dynamical processes to be
described below.

Notice that the action of $d_3$ in Eq.~(\ref{conjb}) is not always
linear! This nonlinearity can be traced to the nonlinearity of the
corresponding supergravity equations of motion. Notice that in the
absence of the $H$-field and NS5 branes, the supergravity equations of
motion, and hence the differential, becomes linear. So what do we mean by
taking the cohomology with respect to $d_3$?  As we will see when we
review interpretation of this sequence due to Maldacena, Moore, and
Seiberg (MMS), we mean simply that physical flux configurations are
those annihilated by $d_3$.  In the linear case we quotient by the
image of $d_3$ because the corresponding states may decay via physical
process.  In fact there are many interesting examples where $d_3$ is
linear, such as all of the examples studied by MMS.  A weak form of
this conjecture may be stated which only includes this subset of
configurations.  In the present paper we consider a stronger form of
this conjecture, in which we claim that all states which are related
by the addition of any element in the image of $d_3$ are related by a
physical process and thus equivalent in the sense of MMS.  The fact
that this is possible is supported by the fact that in known examples
the image of $d_3$ is still a linear space.  It would be interesting
to find and study a counterexample of this.  Of course when $d_3$ is
nonlinear it is an abuse of language to continue to use to the phrases
``differential operator'' and ``spectral sequence''.

We know that this is not a complete list of differential operators.
In particular there are examples \cite{MMS} that illustrate that anomaly
cancellation can require the operator $d_5$, and new instantons can
allow states to decay which are exact under $d_5$ but not under $d_3$.
We consider some such configurations in detail in Section
\ref{susec}. We leave the completion of this list and its interpretation to a
sequel.


\section{Review of Branes in IIB Supergravity} \label{sugsec}
\subsection{Classical Equations of Motion} \label{sugrasec}
The low energy effective theory of IIB string theory is 10-dimensional
IIB supergravity, which has the following action (here the
self-duality constraint for $G_5$ is imposed by hand):

\group{s}
\begin{equation}
S=-\frac{1}{4\kappa_{10}^2}\int d^{10}x(-g)^{1/2}[\mathcal{L}_{NS}+\mathcal{L}_R+\mathcal{L}_{CS}]
\end{equation}
\vspace{-.3in}
\begin{equation}
\mathcal{L}_{NS}=e^{-2\Phi}(-2R-8\partial_\mu\Phi\partial^\mu\Phi+|H|^2)
\end{equation}
\vspace{-.3in}
\begin{equation}
\mathcal{L}_{R}=|G_1|^2+|G_3|^2+\frac{1}{2}|G_5|^2
\end{equation}
\vspace{-.3in}
\begin{equation}
\mathcal{L}_{CS}=H\wedge C_4\wedge dC_2
\end{equation}
\reseteqn 
where we have defined the gauge invariant fieldstrengths

\begin{equation}
G_3=dC_2+H\wedge C_0\sp
G_5=dC_4+\frac{1}{2}H\wedge C_2-\frac{1}{2}B\wedge dC_2. \label{strength}
\end{equation}
In terms of these fieldstrengths the Bianchi identities can be rewritten
 
\begin{equation}
ddB=dH=0\sp
ddC_{p-1}=d(G_p-H\wedge C_{p-3})=dG_p-H\wedge G_{p-2}=0. \label{bianchi}
\end{equation}
The equations of motion are

\begin{equation}
d*H\sim G_3\wedge G_5\sp
d*dC_{p-1}=d*G_p-H\wedge *G_{p+2}=0. \label{h}
\end{equation}

We will be interested in D$p$-branes, fundamental strings, NS5-branes,
M2-branes and M5-branes, which couple to the connections $C_{p+1}$,
$B$, $B^{dual}$, $C_3$ and $C^{dual}$ respectively.  The corresponding
fieldstrengths are \group{strengths}
\begin{equation}
G_{p+2}=dC_{p+1}-H\wedge C_{p-1}\sp
H=dB\sp *H=d{B^{dual}}
\end{equation}
\vspace{-.3in}
\begin{equation}
G_4=dC_3 \sp *G_4=d(C^{dual})
\end{equation}
\reseteqn 

\subsection{Branes Ending on Branes} \label{bendonb}
Imagine that a D$p$-brane wraps a 3-cycle that supports $k$ units of
$H$-flux. As a D$p$-brane gives rise to a transverse $G_{8-p}$-flux,
$G_{8-p}\wedge H \neq 0$ and the supergravity equation of motion from
Subsec. \ref{sugrasec}
\begin{equation}
H\wedge G_{8-p}=dG_{10-p} \label{sug2}
\end{equation}
implies the existence of nonzero D$(p-2)$-brane current $G_{10-p}$
emanating from the D$p$-brane, transverse to both the 3-cycle
supporting the $H$-flux and any $S^{8-p}$ linking the
D$p$-brane. Integrating both sides oen finds that the current can be
associated with the presence of $k$ units of D($p-2$)-brane charge.
Thus, $k$ D($p-2$)-branes must end on any D$p$-brane wrapping a
3-cycle with $k$ units of flux.

The above arguments can be used to demonstrate that a D$p$-brane
wrapped on a $p$-cycle with $k$ units of $G_p$-flux must be the
endpoint of $k$ fundamental strings.  For the case $p=3$ this is the
S-dual of $k$ D-strings ending on a D3-brane wrapping a three-cycle
with $k$ units of $H$-flux.  For the case $p$=5 it is the S-dual of
D-strings ending on an NS5-brane, a configuration to which we will now
turn our attention.

\begin{figure}[ht]
  \centering \includegraphics[width=4in]{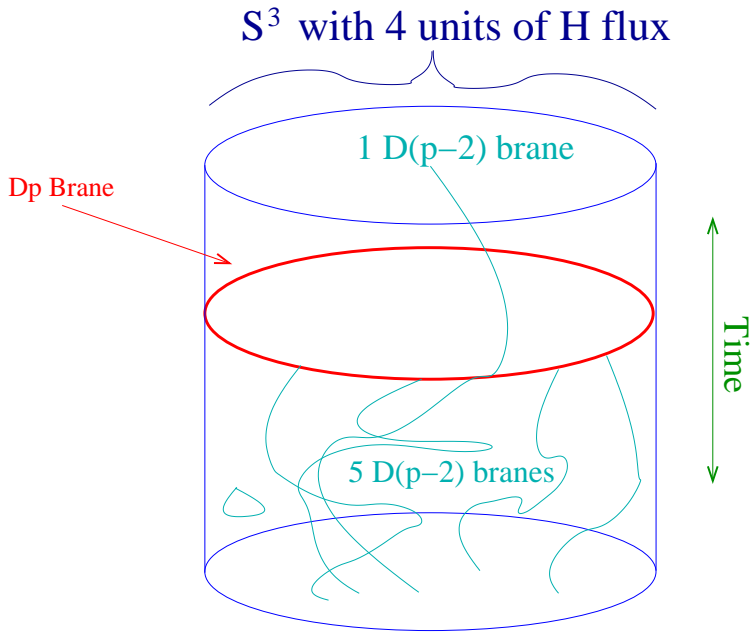}
\caption{A D$p$-brane wraps a 3-sphere that supports 4 units of $H$
flux.  Anomaly cancellation requires that 4 D$(p-2)$-branes end on
this D$p$-brane.  Thus the lone D$p$-brane is not allowed.  Also the
number of D$(p-2)$-branes is only defined modulo 4 because a dynamical
process involving similarly wrapped D$p$-brane instantons can create
or destroy 4 of them at a time.}  \label{mms}
\end{figure}

Consider an NS5-brane that wraps a $p$-cycle $Z$ supporting $k$ units of
$G_{p}$-flux.  Employing the S-dual of the above argument, the NS5
brane is linked by a 3-sphere such that
\begin{equation}
\int_{S^3}H=1.
\end{equation}
Now we may integrate the supergravity equation of motion
\begin{equation}
H\wedge G_p=dG_{p+2}
\end{equation}
over $S^3 \times Z$ to see that there are $k$ D3-branes intersecting
every 3-sphere linking the NS5-brane and so $k$ D3-branes must be
threaded down the NS5-brane's throat.

\section{Maldacena-Moore-Seiberg Construction of the AHSS}
\subsection{Lifting Cohomology to K-Theory}
The K-theory classification of D-branes in IIB string theory
represents all configurations of D-branes as defects in the tachyon
field on the worldvolume theory of a collection of unstable
D9-$\overline{\textup{D}9}$ pairs.  In particular the D$p$-branes are
Poincare dual to $c_{(9-p)/2}$ of an associated vector bundle.  Thus
all D-brane configurations yield a cohomology class, however not every
cohomology class is a Chern class of a vector bundle.

For example, consider a D5-brane which is Poincare dual to some 4-form
$\omega$.  As $\omega$ has no higher form components, the D5-brane
carries no lower brane charges.  Such a D5-brane is allowed if
$\omega$ is $c_2$ of some vector bundle.  In the absence of D7-branes,
this bundle must have
\begin{equation}
c_1=0\sp
c_2=\omega.
\end{equation}
However for any vector bundle
\begin{equation}
c_3=c_1\wedge c_2+Sq^2 c_2\ \textup{mod\ } 2
\end{equation}
where ``mod 2'' means that these two-forms agree as elements of the
$\Z_2$ valued cohomology of spacetime.  $c_1$ vanishes and so $c_3$
must be equal to $Sq^2 c_2$ mod 2.  However a Chern class is an
element of integral cohomology, and so $c_3$ must be the lift of $Sq^2
c_2$ to integral cohomology.  A differential form has such a lift
precisely when it is annihilated by $Sq^1$ and so the existence of the
desired bundle requires
\begin{equation}
0=Sq^1 Sq^2 \omega=Sq^3 \omega
\end{equation}
where the last equality came from an Adem relation.  

Thus we find that a D5-brane configuration corresponds to a section of
some bundle on D9-$\overline{\textup{D}9}$ pairs exactly if its
Poincare dual is annihilated by $Sq^3$.  Thus $Sq^3$ is an obstruction
to lifting an element of $H^4$ to K-theory.

In the presence of an $H$-field we are not interested in K-theory, but
rather in twisted K-theory \cite{WittenK} or perhaps the algebraic
K-theory of sections of a $PU(\infty)$ bundle \cite{BM} or
${\hat{E}}_8$\ bundle. In this case the above obstruction is actually
\begin{equation}
d_3=Sq^3+H.
\end{equation}  
This is one of many obstructions, but remarkably only a finite number
of obstructions exist for a given differential form and these are all
contained in the Atiyah-Hirzebruch spectral sequence (AHSS) reviewed
in Subsec.~\ref{origsec}.  Physically
the failure of this sequence of differentials to annihilate the
Poincare dual of the submanifold wrapped by a D-brane indicates the
presence of an anomaly which in the case $p=3$ is the Freed-Witten
anomaly \cite{FreedWitten}.

Not only do some elements of cohomology fail to lift to K-theory, but
others are equivalent in K-theory.  More precisely, two differential
forms are equivalent as K-theory elements if they differ by a form
which is in the image of any of the above differentials $d_p$.  The
physical interpretation is that equivalent brane configurations are
related by dynamical processes.  In IIA only $p=3$ is nontrivial.  In
IIB $p=3$ is nontrivial and also, when $H\neq 0$, $p=5$ gives a
restriction on allowed D5-brane wrappings and identifies states with
different numbers of D1-branes.

\subsection{The MMS Construction}
Maldacena, Moore, and Seiberg \cite{MMS} propose the following
classification scheme for physical states.  Start with all states
which are consistent time independent backgrounds, in this case that
means all D-branes wrap submanifolds dual to elements of the kernels
of all of the above differential operators.  Then identify states that
are related by physical processes, that is, identify D-branes wrapping
submanifolds whose duals differ by an element of the image of some
differential.  The final result of this classification scheme is
therefore the cohomology with respect to all of the above differential
operators, which in IIB is the associated graded algebra of $K_0$.  If
instead of branes we considere fluxes then we are instead interested
in $K^1$.  After solving an extension problem, one arrives at the
desired K-group.

This prescription has a simple realization in terms of branes ending
on instantonic branes.  Here ``instantonic'' refers either to
solutions of the Euclidean equation of motion giving rise to tunneling
between charge states or solutions of the Lorentzian equations of motion
corresponding to allowed transitions between them.  The idea is that
an instantonic D$p$-brane can wrap a nontrivial 3-cycle which supports
$k$ units of $H$-flux.  Assume that the image of the D$p$-brane is
spin$^c$ or equivalently that its dual is annihilated by $Sq^3$.  As we
saw above, the classical equations of motion require that $k$ D$(p-2)$
branes end on this D$p$-brane.  We will see that this is also required
for anomaly cancellation on the worldsheet theory of fundamental
strings that end on the D$p$-brane.  The result is that the lone D$p$
brane configuration is forbidden and a state consisting of $k$
D$(p-2)$-branes is trivial, as it decays to the vacuum via a process
with an instantonic D$p$-brane.

In the language of the AHSS, the Poincare dual (in the 9-dimensional sense)
$\omega$ of the submanifold wrapped by the D$p$-brane is not
annihilated by the differential $d_3=Sq^3+H$, instead
\begin{equation}
d_3 \omega=\eta
\end{equation}
where $\eta$ is dual (in the 10-dimensional sense) to the submanifold
inhabited by the D$(p-2)$-branes.  Thus the D$p$-brane corresponds to
a form that is not $d_3$-closed and so is forbidden, while the $k$
D$(p-2)$-branes correspond to a form which is $d_3$-exact and so their
configuration is trivial.  In particular they can decay into the
vacuum.

K-theory not only classifies charges, but also classifies fluxes
\cite{MW}.  Technically it is K-cohomology rather than K-homology
\cite{Periwal,HarveyMoore} that classifies fluxes, but this is a
subtlety that we ignore throughout this paper as our goal is simply to
generalize the spectral sequence, and not to learn what generalization
of K-theory our result describes.  The allowed RR fluxes are those
which satisfy
\begin{equation}
d_3 G_p=Sq^3G_p+H\wedge G_p=0. \label{rrcond}
\end{equation}
This is simply the Bianchi identity (\ref{bianchi}) and classical
equation of motion (\ref{h}) in the absence of D-brane sources and with a torsion correction.
Equation (\ref{rrcond}) is precisely the flux version of the statement
discussed above for charges.

\subsection{Generalization}
As suggested in Ref.~\cite{MMS2}, this result can be generalized to
include NS5-branes and fundamental strings.  In particular, an NS5
brane can only wrap a 3-cycle with $k$ units of $G_3$-flux if it is
the endpoint of $k$ D3-branes.  This is the S-dual of the $p=5$ case
of the previous section.  Thus some lone NS5-brane wrappings are not
allowed, while some D3-brane configurations that were nontrivial using
only the considerations of the previous section are actually trivial
because of the dynamical process in which an instantonic NS5-brane
appears from the vaccuum, absorbs the D3-branes and then vanishes
again.

Including the processes of this section and the last, the total number
of D3-branes is conserved only modulo the greatest common divisor of
the $G_3$ and $H$-flux supported on any 3-cycle due to the effects of
instantonic NS5-and D5-branes respectively.  More generally the D3
branes themselves may have nontrivial wrappings, in which case the
relation is more complicated.  These effects may be built into a
generalization of the AHSS in which one begins with both NS and RR
fields and defines $d_1$ to be the exterior derivative in both cases.
Again $d_3$ is defined on RR fluxes by
\begin{equation}
d_3 G_p=(Sq^3+H)G_p. \label{rr1}
\end{equation}
D3-branes ending on instantonic NS5-branes can be incorporated if one defines the action of $d_3$ on $H$ as follows:
\begin{equation}
d_3 H= G_3\wedge H
\end{equation}
where torsion terms will be incorporated in Sec.~\ref{torsec}.

\begin{figure}[ht]
  \centering \includegraphics[width=4in]{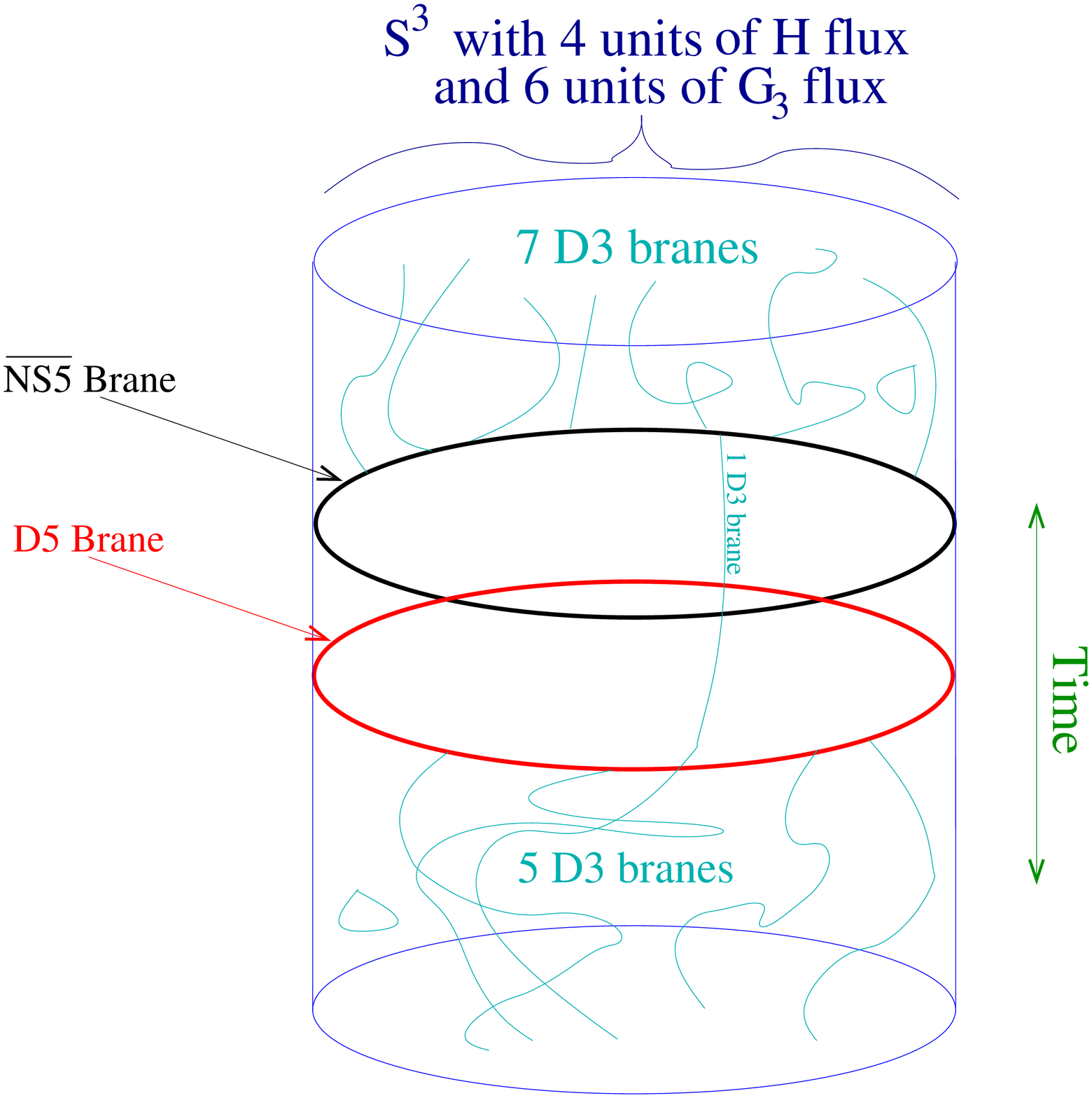}
\caption{A D5-brane and $\overline{\textup{NS5}}$ wrap a three sphere
  that supports 4 units of $H$-flux and 6 of $G_3$-flux.  Anomaly
  cancellation requires that 4 D3-branes end on the D5-brane and 6 begin on
  the $\overline{\textup{NS5}}$.  Therefore one allowed process begins
  with 5 D3-branes, 4 of which decay via the instantonic D5-brane
  leaving just 1.  Later the instantonic $\overline{\textup{NS5}}$
  appears and disappears, leaving 6 more D3-branes for a total of 7.
  Thus we see that the number of D3-branes is only conserved modulo 2,
  where 2 is the greatest common divisor of 4 and 6.}
  \label{d3}
\end{figure}

There is another relevant process.  Recall that a D$p$-brane can wrap
a $p$-cycle which supports $k$ units of $G_p$-flux, in which
case it needs to be the endpoint of $k$ fundamental strings.  Thus
some D$p$-brane wrappings which seemed consistent according to
(\ref{rr1}) actually prove to be inconsistent.  To account for this, one
would like to create the differential $D$ such that
\begin{equation}
DG_p\supset G_{8-p}\wedge G_p \label{gend}
\end{equation}
however in this paper we will restrict our attention to the
generalizations of $d_3$ and $d_5$, which increase the degree of a
form by 3 and 5 respectively.

The degree 3 case of (\ref{gend}) occurs only for the case $p=5$,
corresponding to D3-branes wrapping 3-cycles that support $k$ units of
$G_3$-flux.  These branes must be the endpoints of $k$ fundamental
strings, which is the S-dual of the statement from the previous
section with $p=3$.  Thus the embedding of a D3-brane has two
consistency conditions, one coming from the $H$-flux of a 3-cycle that
it wraps and one from the $G_3$-flux.  $q$ units of $H$-flux and $p$
of $G_3$-flux require that a $(p,q)$ string end on the D3-brane.  This
is summarized by requiring the vanishing of a new differential
$d^\prime_3$ on the $G_5$-flux surrounding a D3-brane:
\begin{equation}
d^\prime_3 G_5=G_3\wedge G_5
\end{equation}
where again torsion will be incorporated in Sec.~\ref{torsec}.

\section{Torsion} \label{torsec}
\subsection{The Freed-Witten Anomaly}
Consider a gauge theory on a manifold $M$ with gauge group $G$.  A fermion that transforms
in the fundamental representation of the gauge group is a section of
the bundle
\begin{equation}
S(M)\otimes G \label{bund}
\end{equation}
where $S(M)$ is a spin ``bundle'' and $G$ is the associated vector ``bundle'' on
which some gauge field $A$ is a connection.  The word ``bundle''
appears in quotes because charged fermions may exist even in the
absence of a spin bundle, that is on a manifold that is not spin.
More precisely, the ``bundle''s in (\ref{bund}) may have transition
functions whose triple products are not the identity, so long as the
triple products in $S(M)$ and $G$ cancel so that the tensor product
(\ref{bund}) is an actual bundle.  Such a bundle defines a spin$^c$
structure on $M$, and so we see that $M$ must be spin$^c$ if the gauge
theory contains charged fermions.

The gauge theory on the worldvolume of a D-brane does contain charged
fermions, arising from the endpoints of fundamental strings.
Therefore D-branes are restricted to wrap spin$^c$ submanifolds of
spacetime which are submanifolds $N\subset M$ whose third
Stieffel-Whitney class vanishes,
\begin{equation}
W_3(N)=0.  \label{w3}
\end{equation} 
The Steenrod square $Sq^3$ is a map from $p$-forms $\omega$ to
$(p+3)$-forms $\eta$ that takes the Poincare dual of $N$ to itself
wedged with $W_3$ of the normal bundle of $N$ pushed forward onto $M$.  This
map is not necessarily an injection and so
\begin{equation}
i_*(W_3(N))\equiv Sq^3(PD(N))=0 \label{sq3}
\end{equation}
is a necessary but not sufficient condition for the satisfaction of
Eq.~(\ref{w3}).  We will see in an example below that combining
(\ref{sq3}) with a degree five differential operator better
approximates the condition (\ref{w3}).

We have just seen that the differential forms Poincare dual to a
configuration of D-branes must be annihilated by $Sq^3$.  Similarly
\cite{MW} the corresponding fieldstrengths must be annihilated by
$Sq^3$.

\subsection{Freed-Witten with $H$-flux}
In the presence of a nontrivial $H$-field the situation becomes
slightly more complicated.  Although the answer appeared in
\cite{WittenK}, the answer was later justified in \cite{FreedWitten}
by an analysis of a global anomaly in the worldsheet path integral of
an open string ending on a D-brane.

In the absence of a $B$-field the path integral measure for a string
with worldsheet $\Sigma$ contains the terms
\begin{equation}
\textup{pfaff}(D)exp(i\oint_{\partial\Sigma}A) \label{nob}
\end{equation}
where $D$ is the worldsheet Dirac operator and $A$ is the gauge
potential of the worldvolume $U(1)$ gauge theory on the D-brane.  The
authors showed that it suffices to consider the case in which the same
spin structure is used for left movers and right movers.  In this case
the Dirac operator is real and so pfaff($D$) must be real.

The pfaffian is real, but there is no natural way to determine its
sign.  Instead one may try to choose a sign, but there may be no
consistent sign choice.  That is, it may be that whatever sign one
chooses, if the worldsheet slides along a particular circle and
returns to its original position the sign may flip.  In fact the
authors proved that the sign flips
\begin{equation}
\alpha=\int_{S}w_2(N)
\end{equation}
times, where $S$ is the surface traced out by $\partial\Sigma$ and
$w_2(N)$ is the second Stieffel-Whitney class of the submanifold into
which the brane is embedded.

Thus the path integral measure is only well defined if the holonomy
$exp(i\oint A)$ changes signs $\alpha$ times as well.  In this case
$A$ could not actually be the connection on a bundle, as this would
imply that $\alpha=0$, but rather $A$ is the connection on a
``bundle''.  The fact that both terms in (\ref{nob}) yield cancelling
contributions implies that the tensor product of the spin ``bundle''
and the ``bundle'' on which $A$ is a connection gives an actual
bundle.  Just as in the last subsection, this bundle provides a
spin$^c$ structure.  The worldvolume and worldsheet approaches in
these two subsections had to agree, as charged fermions on the D-brane
worldvolume are the result of such worldsheets.  In particular, such
an $A$ exists precisely when $N$ is spin$^c$ or equivalently
$W_3(N)=0$.

In the presence of a B-field (\ref{nob}) becomes
\begin{equation}
\textup{pfaff}(D)exp(i\oint_{\Sigma}B)exp(i\oint_{\partial\Sigma}A) .
\end{equation}
Now anomaly cancellation requires that $A$ be chosen so that the
change in holonomy $exp(i\oint A)$ precisely cancels the change in the
product of the first two terms.  The obstruction to the existence of
such an $A$ is no longer simply $W_3(N)$, but there is a new
correction arising from $H=dB$.  The holonomy now needs to change
signs $\int(w_2+B)$ times and so for a general $H$ the condition for
the existence of $A$ is now
\begin{equation}
\beta(w_2+B)=W_3+H=0. \label{splush}
\end{equation}
Thus a D-brane wraps a spin$^c$ submanifold $N$ if and only if $N$
carries trivial $H$, however a D-brane can instead wrap a submanifold
$N$ which is not spin$^c$ in the presence of a nonvanishing $H$ which
is precisely equal to $W_3(N)$.  In particular $2H=0$ on any wrapped submanifold.

\subsection{Strings and NS5-Branes}
We have seen that D-branes can only wrap submanifolds that are
spin$^c$, or more generally, that satisfy (\ref{splush}).  This
analysis is readily extended to fundamental strings and NS5-branes.
First, the restriction is trivial in the case of fundamental strings
as they sweep out two dimensional surfaces, which are automatically spin$^c$.
There is no analog of the $H$ term because there are no objects that
end on fundamental strings in the weak coupling limit of type II.

While we do not know how to extend this argument to the NS5-branes of
IIA, in IIB NS5-branes host worldvolume 5+1 dimensional $U(1)$ gauge
theories.  These theories have charged fermions arising from the ends
of D-strings.  Thus we expect NS5-branes to wrap spin$^c$
submanifolds.  This was seen in an example in Ref.~\cite{WittenB} and in more generality in Ref.~\cite{WittenD}.  However the worldvolume of the D-strings couples to the
2-form $C_2$ and so if this $C_2$ is nontrivial then one may expect
that worldsheet anomaly cancellation on the D-string provides a
correction to $(\ref{nob})$.  In particular if one trusts the S-dual
of the argument above one may suspect that
\begin{equation}
W_3+G_3=0 
\end{equation} 
on the worldvolume of the NS5-brane.

Applying this relation to the $H$-flux whose source\footnote{The
  arguments of Ref.~\cite{MW} suggest that such relations on sources
  also apply to the fields that they create.} was the NS5-brane, one
would find
\begin{equation}
(Sq^3+G_3)H=0.  \label{splusg}
\end{equation}
As evidence for (\ref{splusg}) notice that in the supergravity
approximation it reduces to $G_3\wedge H=0$, which at the level of cohomology is a supergravity equation of motion (\ref{bianchi}).
Eq.~(\ref{splusg}) will serve only as a motivation for our conjecture.

\subsection{Condition on $G_3$ and $H$}
To motivate the condition Eq.~(\ref{conjb}) on the pair $G_3$ and $H$,
let us review the relevant pieces of our argument.  First we know that
in the supergravity limit torsion corrections can be neglected and so
the supergravity equation of motion
\begin{equation}
G_3\wedge H=0\textup{\ \ when\ $G\wedge G=H\wedge H=0$}
\end{equation}
holds at the level of rational cohomology.  Also we use the fact that there are
fermions charged in a $U(1)$ gauge theory on the worldvolume a single
D5 or NS5-brane with no background fluxes to arrive at
\begin{equation}
W_3=0\textup{\ \ when\ $H=0$}
\end{equation}
on a D5-brane worldvolume (the Freed-Witten anomaly) and
\begin{equation}
W_3=0\textup{\ \ when\ $G=0$}
\end{equation}
on an NS5-brane worldvolume.  These imply that $Sq^3$ annihilates the
Poincare duals of the worldvolumes.  Following the reasoning of \cite{MW}
this translates into restrictions of the corresponding fieldstrengths:
\begin{equation}
Sq^3(G_3)=0\textup{\ \ when\ $H=0$}\sp
Sq^3(H)=0\textup{\ \ when\ $G_3=0$}\sp
\end{equation}
respectively.

Furthermore we know \cite{FreedWitten} that a D5-brane can be wrapped on a
manifold with $W_3\neq 0$ when there is a background $H$ that
precisely cancels this $W_3$.  This suggests that the generalization
of the above formulas involves mod 2 additions of Steenrod squares and
fluxes so that they may cancel each other.  There is not enough
information to specify the generalization completely, thus the final
form will be a conjecture.  Neglecting $G_3$ and $H$ independent
terms\footnote{We have no reason to disallow such terms, except that
  they have not been seen in examples.}, there are two natural guesses
for the desired condition.

\noindent
Condition $A$:
\begin{equation}
(Sq^3+H)G_3=(Sq^3+G_3)H=0 \label{conda}
\end{equation}

\noindent
Condition $B$:
\begin{equation}
G\wedge H_3+Sq^3(G_3+H)=0. \label{condb}
\end{equation}

Notice that only condition $B$-is S-duality invariant.  Condition $B$
has also appeared in \cite{DMW} as a possible generalization.  In fact, the
authors verified it via the M-theory partition function in the case of
IIB configuarations that can be obtained via the compactification of
M-theory on a torus.  Below we will provide an example which appears
to exclude\footnote{More precisely it excludes a worldvolume version
  of $A$ in terms of $W_3$'s.} condition $A$ but is consistent with
$B$.

\section{Evidence and SU(3)} \label{susec}
Notice that among all of the restrictions in our conjecture, only
(\ref{conjb}) has a nonvanishing torsion term.  The rest are simply
implications of the supergravity equations of motion.  Thus
Eq.~(\ref{conjb}) deserves the most scrutiny.

Of course there is more to the conjecture than simply the
restrictions, there is also the assertion that physical states which
are $d_3$ exact can dynamically decay to the vacuum via instantonic
branes as a result of the MMS construction.

\subsection{NS5-Brane Backreaction}
String perturbation theory breaks down in the near-horizon region of
an NS5-brane when $g_s$ is small at infinity.  This means that one
requires extra care when formulating arguments concerning the physics
of this region.  Here we present three reasons to trust the arguments presented above.

First, as already noted, in every case but one the restriction
is simply a well-known supergravity equation of motion.  Our only
contribution is to interpret these equations in the MMS framework so
as to show the compatibility of the K-theory framework with S-duality
at the level of $d_3$.

Second, the supergravity arguments are formulated on a sphere
consisting of points at any fixed distance from the NS5-brane, in
particular a distance can be chosen to be large enough so that string perturbation theory
is valid.  The fact that the signed intersection number of D3-branes
with this tube is equal to the $G_3$-flux on a 3-cycle wrapped by the
NS5-brane indicates that the 3-branes must thread down the
throat of the NS5-brane as desired.

Finally, the torsion arguments are the result of the anomaly structure
of the NS5-brane worldvolume theory.  In particular, U(1) charged
fermions require a spin$^c$ structure, or equivalently the Pfaffian of
the Dirac operator must be well defined.  Such arguments tend to be
stable under deformations of the theory.  In particular, one could
increase $g_s$ thereby smoothing the geometry around the NS5-brane and
then make the same arguments, which one expects to still hold when the
asymptotic string coupling is turned back down.  One may still be
concerned that strong coupling effects near the NS5-brane lead to
additional terms in the D-string worldvolume that may cancel this
anomaly, such as the $G_3$ term which can cancel the anomaly in the
case of an NS5-brane in a RR background.  If such terms do exist
then perhaps one may learn about them by understanding when this
anomaly argument fails.

In addition to the near-horizon geometry of NS5-branes, there is also
reason to be concerned about the validity of S-duality, particularly
in cases with less than 16 supercharges and torsion cohomology classes
\cite{AspinwallPlesser}.  For this reason we do not invoke S-duality
in such cases.  Yet our conjecture does turn out to be consistent with
S-duality.

\subsection{(1,1) 5-branes on Non-Spin$^c$ Manifolds}

Consider a $(1,1)$ 5-brane wrapped around a submanifold $N\subset M$
with no background fluxes that restrict nontrivially to the
submanifold.  The
brane will create fluxes $G_3=H$ and so in particular the conditions
on the fluxes (\ref{conda}) and (\ref{condb}) can be reexpressed\\

\noindent
Condition $A$:
\group{condaa}
\begin{equation}
0=Sq^3 G_3+H\wedge G_3=G_3\wedge G_3+G_3\wedge G_3=2 G_3\wedge G_3
\end{equation}
\vspace{-.3in}
\begin{equation}
0=Sq^3 H+H\wedge G_3=H\wedge H+H\wedge H=2 H\wedge H
\end{equation}
\reseteqn

\noindent
Condition $B$:
\begin{equation}
0=Sq^3 (G_3+H)+H\wedge G_3=G_3\wedge G_3+H\wedge H+G_3\wedge H=3G_3\wedge G_3=Sq^3(G_3).
\end{equation}
In particular condition $A$ is always satisfied.  Condition $B$ is not
satisfied precisely when $Sq^3 G_3\neq 0$.  In particular if condition
$B$ is not satisfied then $N$ is not spin$^c$, because $Sq^3 G_3$ is a
pushforward of $W_3(N)$.

However a (1,1) 5-brane carries a U(1) gauge field and has charged
fermions corresponding to (1,1) strings that end on it.  Thus the
worldvolume must have a spin$^c$ structure so that the Dirac operator
for these fermions can be constructed.  This means that in fact there
is an anomaly if a (1,1) 5-brane wraps a non-spin$^c$ submanifold.
And so we learn that whenever $W_3(N)$ is not in the kernel of the
pushforward $i_*$ of the inclusion $i:M\hookrightarrow N$, the anomaly
is predicted by condition $B$ but not predicted by condition $A$,
which is always satisfied for a (1,1) 5-brane with no background
fluxes.  This is yet another piece of evidence in favor of condition
B, which is the condition chosen in our conjecture.

While unfortunately $i_*$ does kill $W_3(N)$ for all submanifolds $N$
that we know how to wrap branes around\footnote{An exception may be
the $\R$P$^5$ in Ref.~\cite{WittenB}, but in our paper, which is about
IIB, we do not consider configurations with orientifolds.}, we will
now illustrate an example of this (1,1) 5-brane anomaly in a case
where $W_3(N)\neq 0$ but $Sq^3=0$.

\subsection{The Topology of SU(3)}

We will consider type II string theory on SU(3)$\times\R^{1,1}$.  We
will begin with type IIA, reviewing the results of \cite{MMS} and
then will employ T-duality to understand the scenario of interest.
First we will describe the relevant features of the topology of the
SU(3) group manifold.

The nonvanishing integral homology classes of SU(3) are
\begin{equation}
H_0(SU(3),\Z)=\Z, ~~ H_3(SU(3),\Z)=\Z, ~~
H_5(SU(3),\Z)=\Z, ~~ H_8(SU(3),\Z)=\Z,
\end{equation}
which are represented by the submanifolds $p$, $S^3$, $M_5$, and
$SU(3)$ respectively, where  $p$ is a point, $S^3$ is an embedded $SU(2)$
and $M_5$ is the group manifold SU(3)/SO(3).  Similarly the integral
cohomology ring is trivial except for $H^0$, $H^3$, $H^5$, and $H^8$
which are isomorphic to $\Z$ and generated by $1,\ x_3,\ x_5$, and
$x_8$ which are Poincare dual to SU(3), $M_5$, $S^3$, and $p$
respectively.  The $\Z_2$ homology and cohomology classes are the same
with $\Z$ replaced by $\Z_2$.

The submanifold $M_5$ is more interesting.  This has nonvanishing
integral homology classes
\begin{equation}
H_0(M_5,\Z)=\Z\sp H_2(M_5,\Z)=\Z_2\sp H_5(M_5,\Z)=\Z
\end{equation}
with generators $q$, $M_2$ and $M_5$ respectively where $M_2 \sub M_5$ is a
2-sphere.  By the universal coefficient theorem the $\Z_2$ homology
is similar
\begin{equation}
H_0(M_5,\Z_2)=\Z_2\sp H_2(M_5,\Z_2)=\Z_2\sp H_3(M_5,\Z_2)=\Z_2\sp H_5(M_5,\Z_2)=\Z_2
\end{equation}
where $H_3(M_5,\Z_2)$ is generated by an $M_3 \sub M_5$ whose boundary wraps the
2-sphere $M_2$ twice.  The integral cohomology classes are
\begin{equation}
H^0(M_5,\Z)=\Z\sp H^3(M_5,\Z)=\Z_2\sp H^5(M_5,\Z)=\Z
\end{equation}
where the generator of $H^3(M_5,\Z)$ is $W_3(M_5)\neq 0$, the third
Stieffel-Whitney class of $M_5$.  In particular $M_5$ is not spin$^c$.
The $\Z_2$ cohomology is
\begin{equation}
H^0(M_5,\Z_2)=\Z_2\sp H^2(M_5,\Z_2)=\Z_2\sp H^3(M_5,\Z_2)=\Z_2\sp H^5(M_5,\Z_2)=\Z_2
\end{equation}
where $H^2(M_5,\Z_2)$ is generated by $W_2(M_5)$, the second
Stieffel-Whitney class.

The authors consider a background $H$-field of $H=k x_3$.  They then
show that this $H$-field restricts nontrivially to $M_5$:
\begin{equation}
x_3|_{M_5}=W_3(M_5).
\end{equation}
They conclude that if a D6-brane wraps $M_5$ precisely once, the
condition for worldvolume anomaly cancellation is the constraint
\begin{equation}
0=W_3(M_5)+H|_{M_5}=(1+k)W_3(M_5)
\end{equation}
which is satisfied precisely when $k$ is odd.

\subsection{A (1,1) 5-brane on $M_5$}
Compactify the noncompact spatial direction on a circle, T-dualize
with respect to it and then take the radius back to infinity.  This
gives the same configuration as above but in IIB.  Now one can wrap a
(1,1) 5-brane around $M_5$.  Notice that (1,1) strings may end on the
5-brane and they will yield fermions charged under the $U(1)$
worldvolume gauge group.  In the absence of any external fluxes that
might affect the path integral of the (1,1) string, anomaly
cancellation requires that the submanifold wrapped by the 5-brane be
spin$^c$.  However $M_5$ is not spin$^c$ and so this configuration is
anomalous.

So we see that $W_3\neq 0$ and there is an example of the kind of
anomaly employed throughout this paper.  Unfortunately $W_3$ is
annihilated by $i_*$ and so $Sq^3=0$, thus the anomalous configuration
is not excluded by $d_3$.  However, as seen in the T-dual situation in
Ref.~\cite{MMS}, the anomaly is detected by $d_5$.

\subsection{The MMS Instanton and $d_5$}
Consider IIB on SU(3)$\times\R^{1,1}$ with a background flux $H=kx_3$.
Following Ref.~\cite{MMS}, anomaly cancelation on the worldvolume of an
instantonic D7-brane wrapped around the SU(3) implies that such
instantons are the endpoint of $k$ D5-branes which each wrap $M_5$.
Thus a state consisting of a multiple of $k$ such D5-branes can
dynamically decay to the vacuum.  Similarly one can wrap an
instantonic D3-brane around $M_3\times\R$ and anomaly cancellation
requires $k$ D1-branes extended along $\R$ to end on it.  Thus D1-brane
number is also, at most, conserved modulo $k$.

To summarize, after taking the cohomology with respect to
$d_3=Sq^3+H$, we find that D5-branes wrapped on $M_5$ are classified
by $\Z_k$ and D1-branes wrapped on $\R$ are also classified by $\Z_k$.
However $M_5$ is not spin$^c$, therefore if a D5-brane wraps $M_5$ $r$
times, the anomaly cancellation condition on the brane worldvolume is
\begin{equation}
0=r(W_3+H)=r(1+k).
\end{equation}
When $r$ is even this is always satisfied and so an even number of
wrappings is always permitted.  However an odd number of wrappings is
anomalous whenever $k$ is even.  Thus these D5-branes are actually
only classified by elements of $2\Z_k=\Z_{k/2}$, in disagreement with
the above calculation obtained using only $d_3$.  Notice that when $H$
and so $k$ vanishes, the D5-branes continue to be classified by $\Z$,
although half of the wrappings that one would expect become anomalous.
Thus one would like to find that $d_5$ is nontrivial precisely when
$k$ is even and $k>0$.

As a hint, we examine instantonic D5-branes wrapped on $M_5\times\R$.
The anomaly cancellation condition is
\begin{equation}
0=W_3(M_5)+H|_{M_5}=(1+k)x_3
\end{equation}
where $x_3$ is the generator of $H^3(M_5,\Z_2)=\Z_2$.  This is
satisfied when $k$ is odd.  When $k$ is even, anomaly cancellation
requires an instantonic D3-brane whose boundary is $M_2\times\R\subset
M_5\times\R$.  Recall that $M_2$ is the generator of
$H_2(M_5,\Z)=\Z_2$ and so it is not a boundary in $M_5$.  However
$H_2(SU(3))=0$ and so $M_2$ must bound a 3-manifold $X$ in SU(3).  The
instantonic D3-brane wraps some $X\times\R$.  We have seen that the
boundary of $M_3$ is two copies of $M_2$ and in fact two copies of $X$
can be deformed to $M_3$.  Moreover the $H$-flux restricted to $X$ is
precisely half of the total $k$ units of $H$-flux.  As $X$ is a 3
manifold, it is trivially spin$^c$ and so anomaly cancellation only
demands that $k/2$ D1-branes extended along $\R$ must end on the
D3-brane.  Therefore this double-instanton violates D1 charge by $k/2$
units, meaning that like D5-branes wrapped on $M_5$, D1-branes are
classified by $\Z_{k/2}$.  Again this disagrees with the result
obtained by simply taking the cohomology with respect to $d_3$.  As
recognized in Ref.~\cite{MMS}, the reason for this discrepency is that
$d_5$ acts via
\begin{equation}
d_5(x_3)=\frac{k}{2}x_5\cup x_3=\frac{k}{2}x_8
\end{equation}
but it would be desirable to have a formula in terms of a
cohomological operation, so that it may generalize to other examples.

\begin{figure}[ht]
  \centering \includegraphics[width=6in]{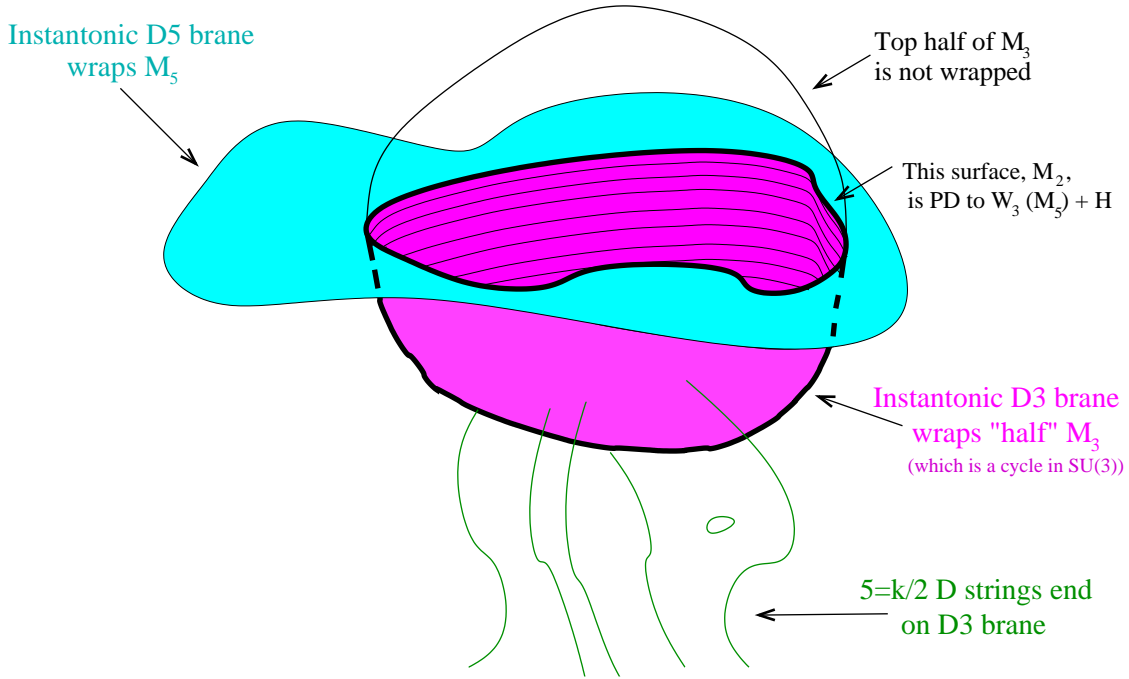}
\caption{The double-instanton of Maldacena, Moore, and Seiberg for
  $k=10$.  The primary instanton is an instantonic D5-brane which
  wraps $M_5\times\R$ while a secondary instantonic D3-brane wraps
  $\R$ crossed with a 3-cycle in SU(3).  This 3-cycle is bounded by
  $M_2$, the Poincare dual of $W_3(M_5) + H$ in $M_5$.  Anomaly
  cancellation on the secondary instanton implies that it must devour
  $k/2=5$ D-strings extended along $\R$.}
  \label{double}
\end{figure}

The crucial observation is that this contribution to $d_5$ is the
result of an instanton with a secondary instantonic brane that is
bounded by $M_2$ which is Poincare dual in $M_5$ to $W_3(M_5)+H$.  The
fact that the boundary is PD to $W_3+H$ is independent of the example,
it was the condition for anomaly cancellation on the primary
instantonic brane.  The secondary instanton wraps ``half of $M_3$'',
which is PD in $M_5$ to $w_2+B$.  Again this fact is example
independent as a result of the following argument.  The primary
instanton's anomaly required that $H^3(M_5,\Z)$ contain a $\Z_2$ class
$W_3+H$.  The fact that this is a $\Z_2$ class implies that there is
some 2-cochain $b$ whose coboundary is precisely twice the 3-cochain
$c$ corresponding to $W_3+H$.  Now consider the dual chain complex and
let $\beta$ be the dual 2-chain to $b$ and $\gamma$ the dual 3-chain
to $c$.  Then $\partial(c)=2b$.  The boundary of the secondary
instanton wraps $b$ and so the rest of the instanton must wrap $c/2$,
an expression which only makes sense when interpreted in terms of the
total spacetime which, being spin$^c$ itself, may be required to have
such a cycle.  If the total spacetime does not have such a cycle, then
this double-instanton cannot exist and thus the contribution to $d_5$
is zero.

To review, the double-instanton consists of a primary instanton which
wraps a submanifold $N$ and a secondary instanton whose boundary wraps
a submanifold of $N$ PD to $W_3(N)+H$.  The worldvolume of the
secondary instanton is $c/2$ where $c$ is the homology class PD to
$w_2(N)+B$ and $c/2$ is interpreted as a chain in the total space.
The secondary instanton itself is subject to anomaly cancellation.
That is
\begin{equation}
W_3(c/2)+H|_{c/2}=0.
\end{equation}
In our example and the T-dual example of Ref.~\cite{MMS}, the dimension of
the secondary instanton is small enough that the $W_3$ term can be
ignored and we will ignore it for now.  It seems quite possible that
this happens in general.  This leaves
\begin{equation}
0=H|_{c/2}=\frac{1}{2}H|_c=\frac{1}{2}(H\cup (w_2(N)+B))|_N. \label{d5world}
\end{equation}
We would like a cohomological formula in terms of the PD of $N$ in the
total space, or in terms of the dual fieldstrength of a D-brane
wrapped on $N$.  To obtain this, we simply pushforward
Eq.~(\ref{d5world}) onto the total space $i:N\hookrightarrow M$.  This
yields the condition
\begin{equation}
\frac{1}{2}i_*(H\cup (w_2(N)+B))=0.
\end{equation}
Thus the contribution of the double-instanton to $d_5$ appears to be
\begin{equation}
d_5(\omega)=\frac{1}{2}i_*(H\cup(w_2(\textup{PD}(\omega))+B)). \label{d5}
\end{equation}
If division by two is not possible, this double-instanton must create
a state with half a unit of D-string charge.  An example of such a
state will be described below.

Notice that in the SU(3)$\times\R^{1,1}$\ case
\begin{equation}
d_5=\frac{1}{2}H\cup Sq^2.
\end{equation}
Like (\ref{d5}), this formula is difficult to decipher.  Applying
$d_5$ to $x_3$ we find the cup product of $H$ with $Sq^2(x_3)$.  This
cup product is well defined in $\Z_2$ cohomology, but we do not want
to restrict $H$ to $\Z_2$ cohomology because we need $k/2$.  Instead we interpret this expression as follows.  Include
$\Z_2=H^5(SU(3),\Z_2)\into H^5(SU(3),\Z)=\Z$ such that the inclusion
is the identity modulo 2.  This is not a homomorphism and it is not
canonical.  Different choices of inclusion will yield different
expressions for $d_5$, but these expressions will differ by $H$ cupped
with a 2-cocycle.  However recall that we have quotiented by $d_3=H$
and so in this quotient ring the different possible expressions for
$d_5$ are all equal and so $d_5$ is well defined, although it is not
well defined on $H^*(SU(3))$.

We do not claim that this is a complete expression for $d_5$, merely
that it is the contribution to $d_5$ from the double-instanton of
Ref.~\cite{MMS}.  In particular, $d_5$ can also contain mod 3 cohomology
operations like $\beta P^2$, which may be calculable from $\Z_3$
phases in M-theory such as those in Ref.~\cite{FluxQuant}.

\subsection{$d_5$ with S-duality and an Example}

We expect that the reinclusion of NS5-branes and background RR flux
will allow an S-duality invariant generalization of the above formula
for $d_5$.  In particular there should be two conditions in such a
generalization, reflecting the requirements that both the number of
fundamental strings and the number of D-strings ending on the double
instanton vanish.  Rather than presenting a general formula, we will
simply investigate the example of IIB string theory on
SU(3)$\times\R^{1,1}$.  Again we will include a background $H$-flux
$H=kx_3$ but now we will also include a background $G$-flux $G=lx_3$.
Let both $k$ and $l$ be nonzero.

As in Ref.~\cite{MMS}, $Sq^3$ acts trivially on the cohomology ring of
SU(3) and so $d_3$ acts simply by multiplication by $G$ or $H$.  This
means that any brane which wraps the $S^3$ in SU(3) is anomalous.  In
particular, an instantonic D3-brane wrapping this $S^3\times\R$ will
be the endpoint of $k$ D-strings and $l$ F-strings.  Thus, neglecting
the effects of $d_5$, D-strings are classified by $\Z_{k}$ and
F-strings by $\Z_{l}$.

A similar classification for 5-branes is more difficult because of our
ignorance about the nature of D7-branes.  Fortunately, such a
classification will not be necessary for the rest of the example.  An
instantonic D7-brane that wraps SU(3) will be the endpoint of $k$ D5
branes and so D5-branes are classified, at the level of $d_3$, by
$\Z_k$.  If one trusts S-duality in such a setting then there will be
a similar instanton involving a 7-brane and $l$ NS5-branes, and so
NS5-branes will be classified by $\Z_l$, ignoring again the fact that
some of these configurations may not turn out to be $d_5$ closed.

Thus far we have obtained an approximate classification using only the
differential $d_3$.  A more thorough classification is obtained by
including the double-instanton.  For concreteness, we will restrict
our attention to $(1,0)$, $(1,1)$, and $(0,1)$ 5-branes. Notice that a
$(1,1)$ 5-brane is only relevant if it gives rise to a single
instantonic D3-brane, but in this case an instantonic D5 and an
instantonic NS5 would have yielded the same D3-brane.  Thus the
$(1,1)$ 5-brane does not need to be considered separately to classify
$(p,q)$ strings in this example.  This leaves us with only two
double-instantons to consider, which have primary instantons that are
a single D5-brane and a single NS5-brane respectively.

Imagine that $k$ and $l$ are both odd.  In this case the induced $H$
and $G$-flux on $H^3(M_5,\Z)=\Z_2$ are each equal to $W_3(M_5)$.  In
particular this means that $W_3+H=W_3+G=0$ on $M_5$ and so a single
D5-brane or NS5-brane instanton is not anomalous and there is no
double instanton.  Thus the 5-branes and strings continue to each by
classified by $\Z_k\times\Z_l$.

Next try $k$ even and $l$ odd.  Thus $W_3+G=0$ on the worldvolume of
an NS5-brane and so if one trusts S-duality\footnote{Our goal in this
subsection is not to test S-duality, but rather to learn what it can
tell us about $d_5$.} then again there is no double-instanton with an
NS5-brane.  Of coures, since $W_3+H \neq 0$ odd numbers of D5-branes are
forbidden, leaving D5-branes to be classified by $\Z_{k/2}$. In
addition, one seems to get a double-instanton consisting of a D5 and a
D3-brane.  Anomaly cancellation on this D3-brane would appear to
require $k/2$ D-strings and $l/2$ F-strings.  However $l/2$ is not an
integer and so this double instanton must be absent. Thus,
unlike the case of $l$ even, D-strings are classified by
$\Z_k$ rather than $\Z_{k/2}$. 

However, this is not the end of the story. Consider a D5-brane
instanton wrapped on $M_5 \times \R$. Cancel the anomaly by adding a
physical D3-brane which extends along $\R$ and is bounded by the PD of
$W_3 + H$ in $M_5$. This D3-brane is wrapped on $X = M_2 \times
\R^{1,1}$, where we recall that this $M_2$ is the $S^2$ which
generates $H_2(M_5,\Z)$ and is also the equator of the $SU(2)$ in
$SU(3)$.  Since $H|_{X} = 0$, anomaly cancellation proceeds with no
strings attached. Instead of an anomaly, the interior of $X$ contains
$k/2$ units of $H$-flux and $l/2$ units of $G_3$-flux. Such branes
could absorb or emit single D-strings and F-strings by expanding or
contracting around one more unit of $G_3$ or $H$-flux, and so the two
quantum numbers of $(p,q)$ strings could be embedded in the two
quantum numbers of these 3-branes supported by flux. One possible
interpretation is that these flux-supported D3-branes carry
half-integral string charge, and require us to augment our initial
chain complex. This is suggestive of a manifestation of
the Myers dielectric effect in our formalism. Further investigation of
this phenomenon might shed light on the analogue of the extension
problem for the extended spectral sequence. On the other hand, notice
that $l/2$ is not an integer, and so we do not know if this
configuration is consistent. If $k$ is odd and $l$ is even the story
is the S-dual of the above case.

The last possibility is that $k$ and $l$ are both even.  Now there are
double-instantons consisting of both an NS5-D3 pair and a D5-D3 pair.
Both instantons violate $(p,q)$ string charge by $(l/2,k/2)$.  Thus $(p,q)$
strings are classified by $(\Z_{k}\times\Z_{l})/\Z_2$.  Notice that
the $(1,1)$ fivebrane yields the same double-instanton.  This is a
manifestation of the nonlinearity of the anomaly cancellation
condition.  In general this perspective does not tell us which $(p,q)$
five branes are anomalous and so does not allow us to classify $(p,q)$
5-branes.

The above results are summarized on the following table:

\vspace{.3in}
\hspace{-.3in}
\begin{tabular}{c|c|c|c|c|c}
k&l&D5-Instanton&NS5 Instanton&$(p,q)$ strings\\ \hline
Odd&Odd&No&No&$\Z_{k}\times\Z_l$\\
Odd&Even&No&Single&$\Z_{k}\times\Z_l$\\
Even&Odd&Single&No&$\Z_{k}\times\Z_l$\\
Even&Even&Double&Double&$(\Z_{k}\times\Z_l)/\Z_2$\\
\multicolumn{4}{c}{}\\
\multicolumn{6}{c}{Table 1: Possible Instantons and String Classification} 
\end{tabular}

\vspace{.3in} Here we have omitted D3-branes wrapping trivial cycles
supported by potentially half-integral fluxes and carrying
$(p,q)$-string charge.
\section{M Theory}

A systematic analysis of the above example may lead to an S-duality
covariant extension of $d_5$, but what variant of K-theory does this
describe?  Perhaps the best source of clues as to the nature of any
such mysterious variant of K-theory is M-theory.  The separation of
fields into NS and RR is a result of the way in which M-theory is
compactified and so an understanding of the classification of these
fields is likely to be a by-product of a classification of fields in
M-theory combined with the action of Kaluza-Klein reduction on this
classification.

In M-theory there is no sense in which we can work at weak coupling.
Nonetheless we will use the MMS prescription to classify M2 and
M5-brane configurations.  In the absence of any understanding of the
quantum theory on these branes we will refrain from discussing the
torsion part of this problem, and instead concern ourselves with the
supergravity approximation.  Distances considered will be much larger
than the 11-dimensional Planck scale.

The 11-dimensional supergravity action contains the terms
\begin{equation}
S\supset \int G_4\wedge\star G_4+C_3\wedge G_4\wedge G_4\sp
G_4=dC_3.
\end{equation}
This leads to the equation of motion
\begin{equation}
d*G_4=G_4\wedge G_4
\end{equation}
which, similarly to the D-brane case, implies that an M5-brane wrapped
around a 4-cycle that supports k units of $G_4$-flux is the endpoint
of $k$ M2-branes.

This suggests that a spectral sequence begins with 
\begin{equation}
E_1=E_1^5\oplus E_1^8=H^5(M,\Z)\oplus H^8(M,\Z) \label{m1}
\end{equation}
and has a single differential
\begin{equation}
d_4=G_4 \label{m2}
\end{equation}
which is clearly trivial on $H^8$ but need not be trivial on $H^5$.

What is this a spectral sequence for?  Following
\cite{FluxQuant,DMW} one can interpret $G_4$ as $p_1$ of an
$E_8$ bundle over $M$.  The above authors considered a bundle not on
the 11-dimensional manifold, but on a 12-dimensional auxillary
manifold that the 11-dimensional manifold bounds.  However the bundle
can be restricted to the 11-dimensional manifold and in fact the
authors proved that the choice of 12 manifold is irrelevant.

$\pi_3(E_8)=\Z$ and all other $\pi_{n<15}(E_8)=0$.  Thus $E_8$ bundles
on manifolds with dimensions of less than 16 are classified by their
first Pontrjagin class $p_1$, reflecting the nontrivial $E_8$ bundles
over $S^4$'s in the 4-skeleton of $M$ where the transition function on
the $S^3$ equator is an element of $\pi_3(E_8)$.  This $p_1$ can be
identified with $G_4$ resulting in the following interpretation of M5
branes\footnote{Recall that $G_4$ is itself equal to $w_4$ of
  the tangent bundle modulo 2.  Thus there is a mod 2 relation between
  the Pontrjagin classes of the 11-dimensional $E_8$ gauge bundle and
  the tangent bundle of the 11-dimensional space, remenicent of the 10
  dimensional condition in heterotic M-theory \cite{HoravaWitten1,
    HoravaWitten2}.  If one interprets the gauge bundle over the end
  of the world as a restriction of the $E_8$ gauge bundle on the 11
  dimensional space to its boundary boundary, then it is possible that
  the 11-dimensional condition, which arises from membrane worldvolume
  anomaly cancellation, restricts to the 10-dimensional condition,
  which arises from a 10-dimensional gravitational anomaly.  Does this
  interpretation have a greater significance?}.

M5-branes are the defects in the $E_8$ bundle such that the restriction
of the bundle to an $S^4$ linking an M5-brane once is the elementary
$E_8$ bundle described in the previous paragraph.  As the other
homotopy classes of E$_8$ vanish, M5-branes will be the only such topological
defects.  M2-branes arise as the electromagnetic dual of the M5-branes.
In particular the existence of M2-branes is necessitated by the above
anomaly for M5-branes wrapped around cycles of nonvanishing $G_4$
flux.  The existence of M2-branes is also required by the
Hanany-Witten transition, which requires an M2-brane to be created
when two M5-branes cross, as follows from the above supergravity
equation of motion.

Alternately the supergravity equation of motion
\begin{equation}
d*G=G\wedge G
\end{equation}
indicates that M2-branes are dual to $p_2$ of the gauge bundle, although for an $E_8$ bundle the relation
\begin{equation}
p_2=p_1\wedge p_1
\end{equation}
reveals that there is no new topology in this characteristic class.

Therefore M2 and M5-branes can be classified by $E_8$ bundles, and
somehow Eqs.~(\ref{m1}) and (\ref{m2}) are the beginning of an analog
of a spectral sequence for a classification of such bundles.  However
we do not know if one should look at a classification of a single $E_8$
bundle, or a construction more like K-theory where one looks at
equivalence classes of pairs of $E_8$ bundles, reflecting annhilation of
pairs of M10 branes.  M10 branes were conjectured to exist in \cite{Lozano}
and as evidenced in \cite{HoravaIIA} they may carry 11-dimensional vector
multiplets and become the unstable D9 sphalerons of IIA after
compactification on a circle.  The above classification scheme may
suggest that each M10 brane must carry 248 vectormultiplets.

\section{Conclusion}
We have constructed rules that allow us to calculate $E_3$ for a given
manifold and even $E_N$ if one can find the higher differentials by
analyzing dynamical processes.  This new sequence appears to classify
RR and NS charged states, modulo higher differentials.  However we
have no geometric interpretation for what this sequence computes.  In
particular, we do not know if this sequence gives the associated
graded algebra of some variant of K-theory and thus, after solving
some extension problem, provides us with a classification of all
states in IIB string theory in terms of a mysterious collection of
bundles.

Notice that we included the D7-brane but not any possible S-duals of
this brane.  This is because we do not know if or how S-duality should
act on a system with 7-branes.  This may be a deficiency in our
proposal.  However, in spaces with ``no compact directions'' 7
brane excitations have an infinite energy backreaction on distant
geometry.  As a result, it is possible that instantonic 7-brane
processes will be infinitely suppressed and therefore in such a limit
we may be able to ignore them.

The MMS interpretation of the Atiyah-Hirzebruch spectral sequence has
allowed us to blindly calculate equivalence classes of stable
configurations modulo dynamical processes.  In particular we can
calculate which fundamental string and D3-brane states are unstable
due to dynamical processes in IIB.  We were also able, with the help
of a generalized Freed-Witten anomaly, to learn which NS5-brane
configurations are unstable.  In M-theory we did the same for M2 and
M5-branes.

However we are unable to provide more than reckless
speculations as to what these equivalence classes may mean.  Although
we are able to generalize the Atiyah-Hirzebruch spectral sequence, we
do not know what mathematical object the new sequence approximates.
Perhaps the most promising hint lies in the mysterious $E_8$ gauge bundle
formalism for M-theory, as all of the fields of IIB are those of
M-theory compactified on a torus and so the answers in IIB are likely
to be those in M-theory with the extra complications arising from this
compactification.

\noindent 
{\bf Acknowledgements}

\noindent
We would like to express our eternal gratitude to A. Adams,
P. Bouwknegt, D. Freed, S. Ganguli, P. Ho$\check{\textup{r}}$ava,
J. Maldacena, H. Murayama, N. Seiberg, S. Shenker, O. de Wolfe and
Y. Zunger for enlightening comments.  The work of UV was supported in
part by the Director, Office of Science, Office of High Energy and
Nuclear Physics, Division of High Energy Physics of the U.S.
Department of Energy under Contract DE-AC03-76SF00098 and in part by
the National Science Foundation under grant PHY-95-14797.  JE lives in
a soundboard box.
\noindent

\bibliographystyle{ieeetr} 
\bibliography{k}

\begin{thebibliography}{10}

\bibitem{MM}
R.~Minasian and G.~Moore, ``K-{T}heory and {R}amond-{R}amond {C}harges,'' {\em
  J. High Energy Phys.}, vol.~{\bf{11}}, p.~002, 1997.
\newblock hep-th/9710230.

\bibitem{WittenK}
E.~Witten, ``D-{B}ranes and {K}-{T}heory,'' {\em JHEP}, vol.~{\bf{9812}}:019,
  1998.
\newblock hep-th/9810188.

\bibitem{Sen}
A.~Sen, ``Tachyon {C}ondensation on the {B}rane-{A}ntibrane {S}ystem,'' {\em
  JHEP}, vol.~{\bf{9808}}:012, 1998.
\newblock hep-th/9805170.

\bibitem{HoravaIIA}
P.~Ho$\check{\textup{r}}$ava, ``Type {IIA} {D}-{B}ranes, {K}-{T}heory, and
  {M}atrix {T}heory,'' {\em ATMP}, vol.~2, p.~1373, 1999.
\newblock hep-th/9812135.

\bibitem{HarveyMoore}
J.~Harvey and G.~Moore, ``Noncommutative {T}achyons and {K}-{T}heory.''
  hep-th/0009030.

\bibitem{AspinwallPlesser}
P.~S. Aspinwall and M.~R. Plesser, ``D-branes, {D}iscrete {T}orsion, and the
  {M}c{K}ay {C}orrespondence.'' hep-th/0009042.

\bibitem{MMS}
J.~Maldacena, G.~Moore, and N.~Seiberg, ``D-{B}rane {I}nstantons and
  {K}-{T}heory {C}harges.'' hep-th/0108100.

\bibitem{AH}
M.~F. Atiyah and F.~Hirzebruch, ``Vector {B}undles and {H}omogeneous
  {S}paces,'' {\em Proc. Symp. Pure Math.}, vol.~{\bf{3}}, p.~53, 1961.

\bibitem{Gimon}
O.~Bergman, E.~Gimon, and S.~Sugimoto, ``Orientifolds, {RR} {T}orsion, and
  {K}-{T}heory.'' hep-th/0103183.

\bibitem{Stong}
R.~Stong, ``Calculation of $\omega_{11}^{spin}({K}({Z},4))$,'' in {\em Unified
  String Theories} (M.~Green and F.~Gross, eds.), 1985 Santa Barbara
  Proceedings, World Scientific, 1986.

\bibitem{FluxQuant}
E.~Witten, ``On {F}lux {Q}uantization in {M}-{T}heory and the {E}ffective
  {A}ction,'' {\em J. Geom. Phys.}, vol.~{\bf{22}}, p.~1, 1997.
\newblock hep-th/9609122.

\bibitem{DMW}
E.~Diaconescu, G.~Moore, and E.~Witten, ``${E}_8$ {G}auge {T}heory, and a
  {D}erivation of {K}-{T}heory from {M}-{T}heory.'' hep-th/0005090.

\bibitem{BM}
P.~Bouwknegt and V.~Mathai, ``D-{B}ranes, {B}-{F}ields, and {T}wisted
  {K}-{T}heory,'' {\em JHEP}, vol.~{\bf{0003}}:007, 2000.
\newblock hep-th/0002023.

\bibitem{FreedWitten}
D.~Freed and E.~Witten, ``Anomalies in {S}tring {T}heory with {D}-{B}ranes.''
  hep-th/9907189.

\bibitem{MW}
G.~Moore and E.~Witten, ``Self-duality, {RR} fields, and {K}-{T}heory.''
  hep-th/9912279.

\bibitem{Periwal}
V.~Periwal, ``D-brane charges and {K}-homology.'' hep-th/0006223.

\bibitem{MMS2}
J.~Maldacena, G.~Moore, and N.~Seiberg, ``D-{B}rane {C}harges in {F}ive-brane
  backgrounds.'' hep-th/0108152.

\bibitem{WittenB}
E.~Witten, ``Baryons {A}nd {B}ranes in {A}nti de {S}itter {S}pace,'' {\em
  JHEP}, vol.~{\bf{9807}}:006, 1998.
\newblock hep-th/9805112.

\bibitem{WittenD}
E.~Witten, ``Duality {R}elations {A}mong {T}opological {E}ffects {I}n {S}tring
  {T}heory,'' {\em JHEP}, vol.~{\bf{0005}}:031, 2000.
\newblock hep-th/9912086.

\bibitem{HoravaWitten1}
P.~Ho$\check{\textup{r}}$ava and E.~Witten, ``Heterotic and {T}ype {I} {S}tring
  {D}ynamics from {E}leven {D}imensions,'' {\em Nucl. Phys. B},
  vol.~{\bf{460}}, no.~506, 1996.
\newblock hep-th/9510209.

\bibitem{HoravaWitten2}
P.~Ho$\check{\textup{r}}$ava and E.~Witten, ``Eleven {D}imensional
  {S}upergravity on a {M}anifold with {B}oundary.'' hep-th/9603142.

\bibitem{Lozano}
L.~Houart and Y.~Lozano, ``Brane {D}escent {R}elations in {M}-theory.''
  hep-th/0001170.

\end{thebibliography}
\end{document}